\documentclass[aps,prd,twocolumn,showpacs,superscriptaddress,preprintnumbers, floatfix,nofootinbib]{revtex4-2}
\usepackage{amsmath,amssymb}
\usepackage{epsfig}
\usepackage{hyperref}
\usepackage{color}
\usepackage{slashed}
\usepackage{placeins}

\usepackage{amsfonts}
\usepackage{graphicx, rotating}
\usepackage{epstopdf}
\usepackage{latexsym}
\usepackage{rotating}
\usepackage{braket}

\usepackage[export]{adjustbox}
\usepackage[font={small}]{caption}   

\usepackage[starfontserif]{starfont}

\newcommand{\Ocal}{{\mathcal O}}

\definecolor{myred}{rgb}{0.7, 0, 0}
\definecolor{myblue}{rgb}{0, 0, 0.7}
\definecolor{mygreen}{rgb}{0.04, 0.7, 0.5}
\definecolor{mygray}{rgb}{0.1, 0.1, 0.1}

\definecolor{oleg}{RGB}{0, 153, 76}

\hypersetup{colorlinks,citecolor=myred,linkcolor=myblue,urlcolor=myblue,linktocpage=true}

\def\be   {\begin{equation}}   \def\ee   {\end{equation}}
\def\ba   {\begin{array}}      \def\ea   {\end{array}}
\def\bea  {\begin{eqnarray}}   \def\eea  {\end{eqnarray}}
\def\bean {\begin{eqnarray*}}  \def\eean {\end{eqnarray*}}

\def\bry{\begin{array}}
	\def\ery{\end{array}}

\newcommand{\skipnew}[1]{}

\def\mainz{\small{Johannes Gutenberg-Universit\"at Mainz, 55128 Mainz, Germany}}
\def\ucb{\small{Department of Physics, University of California, Berkeley, California 94720, USA}}

\def\HIM{\small{Helmholtz Institute Mainz, 55099 Mainz, Germany}}

\def\GSI{GSI Helmholtzzentrum für Schwerionenforschung GmbH, 64291 Darmstadt, Germany}

\def\UD{\small{Department of Physics and Astronomy, University of Delaware, Newark, Delaware 19716, USA}}
\def\JQI{\small{Joint Quantum Institute, National Institute of Standards and Technology and the University of Maryland, Gaithersburg, Maryland 20742, USA}}

\def\SU{\small{The Oskar Klein Centre, Department of Physics, Stockholm University, 10691 Stockholm, Sweden}}

\begin{document}

\date{\today}
\title{\Large\bfseries Search for fast-oscillating fundamental constants with space missions}

%


\author{Dmitry Budker}
\email{budker@uni-mainz.de}
\affiliation{\mainz}
\affiliation{\HIM}
\affiliation{\GSI}
\affiliation{\ucb}

\author{Joshua Eby}
\email{joshua.eby@fysik.su.se}
\affiliation{\SU}




\author{Marianna Safronova}
\email{msafrono@UDel.edu}
\affiliation{\UD}
\affiliation{\JQI}

\author{Oleg Tretiak}
\email{oleg.tretiak@uni-mainz.de}
\affiliation{\mainz}
\affiliation{\HIM}
\affiliation{\GSI}

\begin{abstract}
While it is possible to estimate the dark matter density at the Sun distance from  the galactic center, this does not give information on actual dark matter density in the Solar system. There can be considerable local enhancement of dark matter density in the vicinity of gravitating centers, including the Sun, the Earth, as well as other planets in the solar system. Generic mechanisms for the formation of such halos were recently elucidated. In this work, we studies the possible halo dark matter overdensities  and corresponding dark matter masses allowed for various objects in the solar system.  We explore spacecraft missions to detect such halos  with instruments such as quantum clocks, atomic and molecular spectrometers designed to search for fast (tens of hertz to gigahertz) oscillations of fundamental constants, highly sensitive comagnetometers, and other quantum sensors and sensor networks. 

\end{abstract}

\maketitle

\section{Introduction}

The nature and composition of dark matter (DM) remains a major unsolved problem in modern physics \cite{Battat2024}. The absence of unambiguous non-gravitational detection of DM of any kind despite decades of intense experimental effort is stimulating theorists to come up with various new scenarios, in turn opening new directions in experimental searches. A well-motivated paradigm that has attracted considerable attention in the last decade is that of ultralight bosonic galactic dark matter \cite{Jackson_kimball_search_2023}, with underlying bosons with mass $<10\,$eV forming a field oscillating at the corresponding Compton frequency. The spin of the boson and its intrinsic parity govern possible non-gravitational interactions the DM field may have with fermions. Specifically, scalar (spin-0, even parity) DM leads to apparent oscillation of fundamental constants; see \cite{Antypas2020ADP} and references therein, while pseudoscalar (spin-0, odd parity) DM produces several kinds of spin-dependent interactions that are first order in the DM field \cite{graham_new_2013,stadnik_axion-induced_2014}.

Although in the so-called standard halo model \cite{2018halo,Evans:2018bqy,Asgari:2023mej}, DM is virialized in the galaxy, it was realized that there can be considerable local enhancement of DM density in the vicinity of gravitating centers, including the Sun, Earth, as well as other planets in the solar system \cite{Banerjee:2019epw,Banerjee:2019xuy,Budker2023_Formation}. While initially it was not clear how such DM halos could be formed, generic mechanisms for such processes were pointed out in recent work \cite{Budker2023_Formation}.

The possible existence of such  halos, gravitationaly bound to Solar system objects, opens an intriguing possibility of directly probing these halos with instruments on board spacecraft such as, for example, the next generation of solar probes with optical atomic or nuclear clocks \cite{Tsai2023_Direct_NatAstr}. An interesting feature of the halos is that the density enhancements peak at different dark matter mass ranges for different Solar system objects. 

In this paper, we explore spacecraft searches with other instruments such as atomic \cite{Antypas2019WRESL,Tretiak2022_RFDMPRL} and molecular \cite{Oswald:2021vtc,Allcock2024spud} spectrometers designed to search for fast (tens of hertz to gigahertz) oscillations of fundamental constants, as well as comparisons of atomic transition frequencies with those of a quartz oscillator \cite{Zhang2023_DM_RF}. These technologies have the advantage of simpler setup and correspondingly easier pathway towards space applications. We find that exploration of different halos leads to drastic increase of the dark matter mass range that can be probed with halo searches. Various quantum technologies described in this work are sensitive to dark matter in different mass ranges, enabling maximizing the discovery potential of space-based quantum dark matter searches.

\section{Dark Matter in the Solar System}
\label{sec:DMSS}

As a starting point, we consider a scalar field $\phi$ with small mass $m_\phi \lesssim {\rm eV}$, described by a Lagrangian of the form
\begin{equation} \label{eq:Lagrangian}
    \mathcal{L}_\phi = \frac{m_\phi^2}{2}\phi^2 - \frac{m_\phi^2}{f_a^2\,4!}\phi^4\,,
\end{equation}
where $f_a\gg {\rm GeV}$ is a high-energy scale of the underlying theory (for axion theories this is the `axion decay constant'). The second term in Eq.\,\eqref{eq:Lagrangian} mediates the self-interactions of the field with an effective dimensionless coupling $-m_\phi^2/f_a^2$, where the minus sign indicates an attractive interaction. Such a field can attain a significant energy density in the early universe and, if stable, can constitute some fraction of the DM content of the universe today~\cite{Preskill:1982cy,Abbott:1982af,Dine:1982ah}.

\begin{table*}
    \centering
    \begin{tabular}{c c c | ccccc | cc}
        Object (label) & $M\,{\rm [kg]}$ & $R\,[10^6\,{\rm m]}$ 
        & $r$ & $\dfrac{\rho_{\rm max}^{\rm peak}}{\rho_{\rm local}}$ 
        & $m_\phi^{\rm peak}\,{\rm [eV]}$ & $f_\phi^{\rm peak}$ & $\tau_{\rm coh}^{\rm peak}$ [s] 
        & Constraint \\
         \hline 
        Sun ($\odot$) & $2\times 10^{30}$ & $700$ & 
        AU & $10^5$ &  $2\cdot 10^{-14}$ & $4.8\,$Hz & $9\cdot 10^6$ &  
        $\rho \lesssim 10^5\rho_{\rm local}$, Ephemerides~\cite{Gron:1995rn, Anderson:1995dw,Pitjev:2013sfa,Tsai:2022jnv} \\
        &&& $0.1$ AU & $10^{18}$ & $10^{-13}$ & $24\,{\rm Hz}$& $7\cdot 10^4$ & $M_\star < 0.05M_\odot$ only \\
        Jupiter (J) & $1.9\times10^{27}$ & $71$ & 
        $R_J$ & $10^{22}$ & $2\cdot 10^{-11}$ & $4.8\,{\rm kHz}$ & $10^4$ & $M_\star < 0.05 M_J$ only \\
        Earth ($\oplus$) & $6\times10^{24}$ & $6.4$ & 
        $R_\oplus$ & $10^{19}$ &  $10^{-8}$ & $2.4\,{\rm MHz}$ & $8$ & $M_\star(R_{\rm LAGEOS}<r<R_{\rm m}) \lesssim 4\cdot 10^{-9}M_\oplus$~\cite{Adler:2008ky} \\
        Moon (m) & $7.3\times10^{22}$ & $1.7$ & 
        $R_{\rm m}$ & $10^{18}$ &  $2\cdot 10^{-8}$ & $4.8\,{\rm MHz}$ & $7\cdot10^3$ &  $M_\star(R_{\rm LAGEOS}<r<R_{\rm m}) \lesssim 4\cdot 10^{-9}M_\oplus$~\cite{Adler:2008ky} \\
    \end{tabular}
    \caption{Summary of objects and their companion gravitational atoms studied in this work. The first three columns characterize the central object: their name, label, mass $M$, and radius $R$. 
    The four central columns characterize the gravitational atom at a distance $r$ from its center, including the mass $m_\phi^{\rm peak}$, Compton frequency $f_\phi^{\rm peak}$, and coherence time $\tau_{\rm coh}^{\rm peak}$ where the density obtains its largest possible value, $\rho_{\rm max}^{\rm peak}$ (c.f. Fig.\,\ref{fig:GAdensity}) 
    The last column lists the strongest observational constraint on each case.
    Note that in all cases, in addition to observational constraints, we impose the condition that the total halo mass $M_\star$ can be no more than $5\%$ of the mass of the object. }
    \label{tab:objects}
\end{table*}

In the standard halo model (see, for example, \cite{Evans:2018bqy,Asgari:2023mej}), the local properties of the DM field are determined by large-scale (larger than $100$s of parsecs) observations in the galaxy. The resulting DM energy density $\rho_{\rm local}\simeq0.4\,{\rm GeV/cm}^3$~\cite{Bovy:2012tw,Read:2014qva} and DM speed $v_{\rm dm} \simeq 10^{-3}$~\cite{Schoenrich:2009bx,Eilers_2019} are key inputs to a large number of terrestrial searches for DM. However, dynamics in the DM sector can significantly modify these predictions on smaller scales, giving rise to, for example, overdensities at the scale of astronomical units (AU) or smaller. Although model dependent, such scenarios can provide new opportunities to discover or constrain DM properties.
Many scenarios of beyond-standard model physics that give rise to small-scale DM structure inside the solar system have been proposed in the literature, including axion miniclusters~\cite{Hogan:1988mp,Kolb:1993zz,Kolb:1994fi}, boson stars~\cite{Kaup:1968zz,Ruffini:1969qy,BREIT1984329,Colpi:1986ye}, and solar basins~\cite{VanTilburg:2020jvl,Lasenby:2020goo,DeRocco:2022jyq}.

In this work, we focus on scenarios in which a significant density of scalar field $\phi$ is gravitationally bound to astrophysical objects of mass $M$ and radius $R$. Due to the effective $1/r$ gravitational potential of the central object, these configurations are often called \emph{gravitational atoms}. 
A small value of $m_\phi \ll {\rm eV}$ implies that the gravitational Bohr radius of $\phi$ in the presence of an external mass $M$ can be astrophysically large,
\begin{equation}
    R_\star \equiv \frac{1}{m_\phi \alpha} \simeq {\rm AU}\left(\frac{1.3\cdot10^{-14}\,{\rm eV}}{m_\phi}\right)^2\left(\frac{M_\odot}{M}\right)\,,
\end{equation}
where $\alpha \equiv G M m_\phi$ is the gravitational coupling, and we have illustrated the solar case with $M_\odot$ the mass of the Sun. The energy density of the gravitational atom is given by
\begin{equation}
    \rho(r) = \frac{M_\star}{\pi R_\star^3} \exp\left(-\frac{2r}{R_\star}\right)\,,
\end{equation}
where $M_\star$ is its total mass.\footnote{Here $\rho(r)$ is normalized such that $4\pi\int_0^\infty \rho(r) r^2 dr = M_\star$.} The stability of a gravitational atom is assured as long as $\rho \lesssim \rho_c$, where
\begin{equation} \label{eq:rhoc}
    \rho_c \equiv 16\alpha^2 
            f_a^2 m_\phi^2\,.
\end{equation}
Above this density the self-interactions are sufficiently strong to cause gravitational collapse~\cite{Chavanis:2011zi,Chavanis:2011zm,Eby:2014fya}. The collapsing gravitational atom is expected to closely mirror the case of collapsing boson stars~\cite{Eby:2016cnq,Levkov:2016rkk,Helfer:2016ljl,Eby:2017xrr,Michel:2018nzt}, whose rapid emission of relativistic scalars gives rise to novel signals in direct detection~\cite{Eby:2021ece,Arakawa:2023gyq,Arakawa:2024lqr,Eby:2024mhd} and cosmology~\cite{Du:2023jxh,Escudero:2023vgv,Fox:2023xgx}. Further exploration of such signals is interesting but lies beyond the scope of the present work.

The case of a gravitational atom bound to the Sun and Earth was previously analyzed in the context of terrestrial direct detection experiments~\cite{Banerjee:2019epw,Banerjee:2020kww}.
Recently it was shown that the self-interactions of the field (the second term in Eq.\,\eqref{eq:Lagrangian}) can lead to direct capture of ultralight bosonic dark matter (UBDM) 
around the Sun from the DM background, leading to overdensities of many orders of magnitude~\cite{Budker2023_Formation}. This capture mechanism is general and therefore can be applied to other objects. However, significant overdensities are only captured when $M \gtrsim v_{\rm dm}/(2\pi G m_\phi)$, and $R_\star \gtrsim R$ is also required so that the $1/r$ approximation holds. Since the former condition favors larger $m_\phi$ and the latter prefers smaller $m_\phi$, this exponential growth is obtained outside the surface of some bodies (including Jupiter and the Sun), but not for others (including the Earth and moon); see Appendix~\ref{app:capture} for details.

More generally, direct observations, for example, solar system ephemerides~\cite{Gron:1995rn, Anderson:1995dw,Pitjev:2013sfa,Tsai:2022jnv} or measurements of local Earth satellites~\cite{Adler:2008ky}, can constrain overdensities of DM in the solar system. We review these constraints in Appendix\,\ref{app:constraints}. 
In Fig.\,\ref{fig:GAdensity}, we illustrate the maximum density $\rho_{\rm max}$ allowed by the constraints above\footnote{Note that in all cases we require $M_\star \lesssim 0.05M$ as a fixed constraint in addition to the observational constraints discussed in Appendix~\ref{app:constraints}. This is a sort of common-sense limit based on the observation that replacing baryonic matter with feebly-interacting dark matter inside the bulk of these objects should, at some level, lead to observable effects. The precise value we choose here is motivated by Ref.~\cite{Kardashev:2005}, who derive a constraint on the non-baryonic mass of the Sun of roughly $0.05M_\odot$, obtained by modeling solar evolution.} as a function of $m_\phi$ for the Sun (orange), Jupiter (red), Earth (blue), and Earth's moon (green), evaluated at a distance $r$ from the center of each object as labeled.
For the Sun and Jupiter, we also show the maximum density that can be obtained by the capture mechanism of~\cite{Budker2023_Formation} (dashed lines). For a moon-bound gravitational atom, we also illustrate the point at which the size of the gravitational atom $R_\star^{(\rm moon)}$ becomes equal to the Roche radius\footnote{The Roche radius is the point at which a body (here, the gravitational atom) would be significantly distorted by the tidal force from the gravitational interaction with a second body (here, the Earth).} $R_{\rm Roche}$ of the Earth (faint green line); for $m_\phi$ below this line, the cloud will be significantly distorted by the gravitational field of the Earth.

Another critical property of UBDM is the timescale for coherent oscillations, also known as the \emph{coherence time}. Its precise value depends on the nature of the occupied states in the gravitational atom (see~\cite{Banerjee:2019xuy,Budker2023_Formation} for detailed discussion). If, for example, the is a $\Ocal(1)$ admixture of excited states to the ground state, this leads to decoherence time $t\simeq \tau_{\rm coh}$, where
\begin{equation} \label{eq:taucoh}
    \tau_{\rm coh} \simeq \frac{2\pi}{m_\phi v_g^2}\,,
\end{equation}
with the effective velocity dispersion $v_g \simeq \sqrt{G M/R_\star} = \alpha$. Note that $\tau_{\rm coh}$ represents a minimum coherence time the gravitational atom might obtain, though the precise value will depend on the formation mechanism. As we will see, even this lower limit on the coherence timescale can be extremely long, and as a result these coherent oscillations can be utilized to improve the sensitivity of UBDM searches we propose.

In Table\,\ref{tab:objects}, we summarize the key properties of each object we consider (name, mass, and radius), as well as several properties of their possible gravitational atoms, including the peak density $\rho_{\rm max}^{\rm peak}$, particle mass as the peak density $m_\phi^{\rm peak}$, and corresponding Compton frequency $f_\phi \simeq m_\phi/(2\pi)$. We also include the estimated (lower limit for the) coherence time $\tau_{\rm coh}$ at the peak parameters; see Eq.\,\eqref{eq:taucoh}. For each object we evaluate the density at the surface of the object, except for the Sun where we consider both AU and $0.1\,{\rm AU}$ distances. Finally, we cite the strongest observational constraint in each case.

\begin{figure*}
    \centering
    \includegraphics[scale=1] {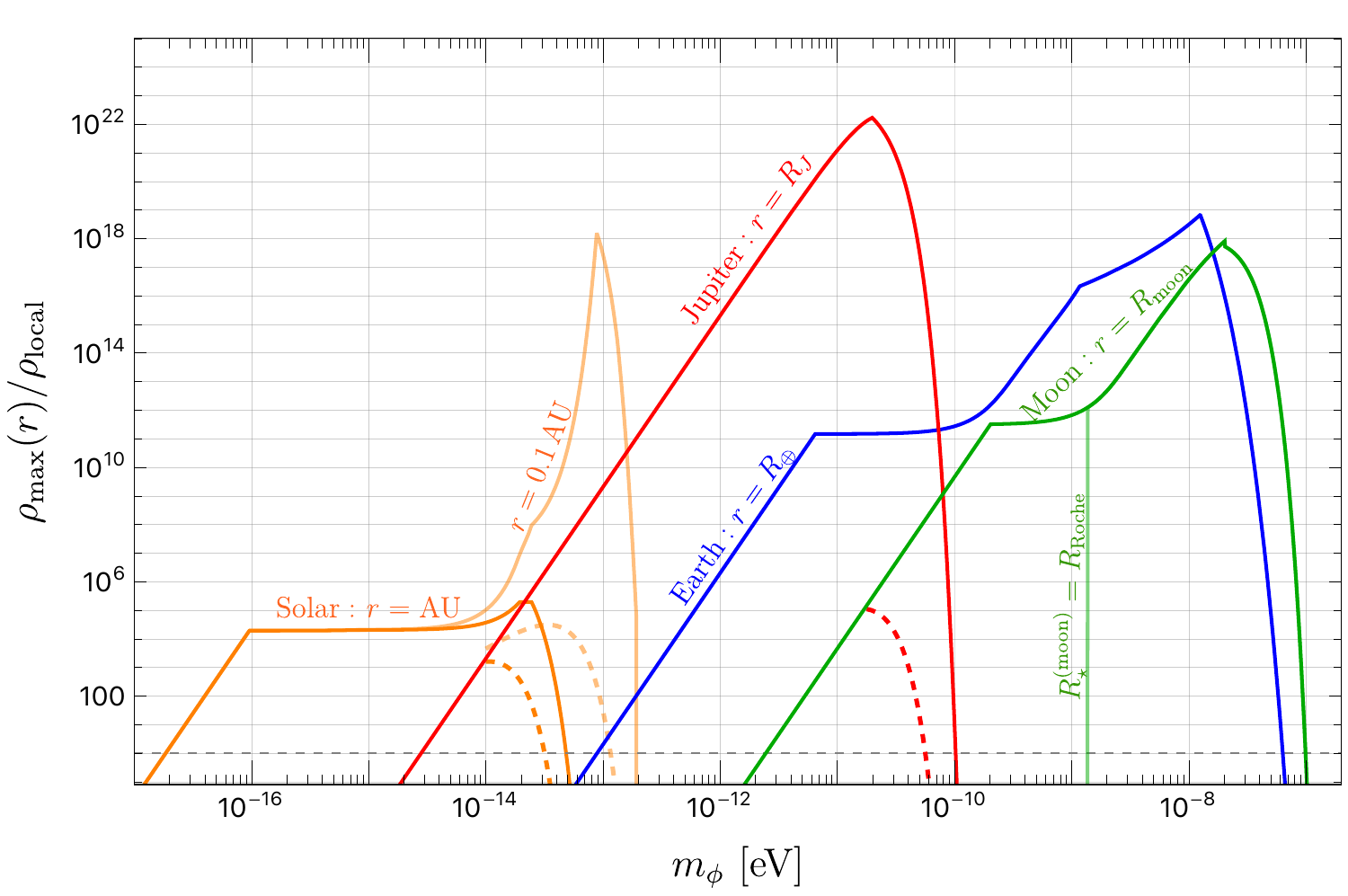}
    \caption{Maximum gravitational atom density as a function of $m_\phi$ bound to the Sun (orange), Jupiter (red), Earth (blue) or moon (green). In each case the density is evaluated at a distance $r$ from the center of the object as labeled. The dashed lines are the maximum density accumulated over $4.5\,{\rm Gyr}$ through the capture mechanism of~\cite{Budker2023_Formation}.
    }
    \label{fig:GAdensity}
\end{figure*}

In this work, we consider the UBDM 
bound to a range of objects in our solar system: the Sun, Jupiter, Earth, and the moon. Though these objects were chosen to characterize the range of masses $m_\phi$ (or equivalently frequencies $f_\phi$) of interest to future missions, they do not exhaust the possibilities. The masses of Mercury, Venus, and Mars, lying between those of Earth and Moon, suggest that their gravitational atoms would be similar to, though in-between, the blue and green curves in Fig.\,\ref{fig:GAdensity}, with the important difference that the local satellite constraints of~\cite{Adler:2008ky} would not apply there. Similarly, a gravitational atom bound to Uranus, Saturn, and Neptune would be similar to those of Earth and Jupiter, lying between the blue and red curves in Fig.\,\ref{fig:GAdensity}. Future missions to these planets could also provide novel opportunities to study the capture of UBDM and possibly lead to discovery.

\begin{figure*}
    \centering
    \includegraphics[scale=0.8] {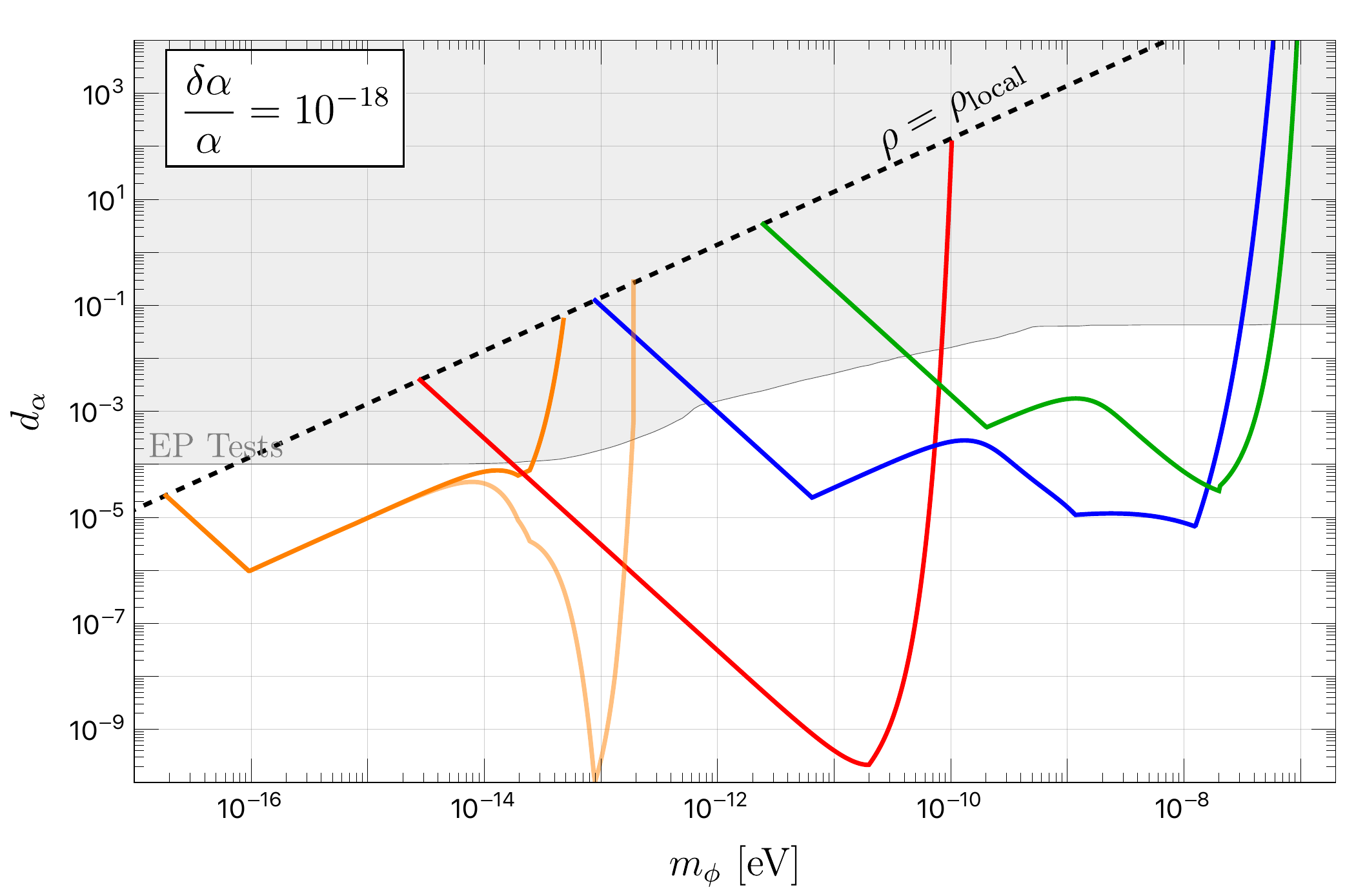}
    \caption{Estimated sensitivity to UBDM-photon coupling $d_\alpha$ and mass $m_\phi$, assuming an fractional sensitivity $\delta\alpha/\alpha=10^{-18}$ (see text for details), for a maximal-density gravitational atom bound to the Sun (orange), Jupiter (red), Earth (blue), or moon (green); the distances are the same as those illustrated in Fig.\,\ref{fig:GAdensity}. Lines are cut by the dashed black line, where $\rho=\rho_{\rm local}$, and the gray shaded region is constrained by searches for violation of the equivalence principle~\cite{Berge:2017ovy,Hees:2018fpg}.
    }
    \label{fig:sensitivity}
\end{figure*}

\section{Searches for Fast-Oscillating Fundamental Constants}
\label{sec:searches}

As illustrated by Fig.\,\ref{fig:GAdensity},  gravitational atom density peaks at different mass ranges for various objects in the solar system. 
The peak masses of dark matter that can form the solar halo are near 10$^{-13}$\,eV.  
A clock-comparison satellite mission with
two quantum (i.e. atomic, molecular, or nuclear) clocks onboard to the inner reaches of the solar system was proposed in \cite{Tsai2023_Direct_NatAstr}
to search for the dark matter halo bound to the Sun and to look for the
spatial variation of the fundamental constants associated
with a change in the gravitation potential. The
projected sensitivity of space-based quantum clocks for detection of Sun-bound dark matter halo exceeds the reach
of Earth-based clocks by orders of magnitude.
Atomic clocks have exquisite fractional sensitivity; however, they operate at relatively low cadence, typically, providing a frequency reading less than a few times per second. In a recent work, a solution was proposed using a broadband dynamic decoupling algorithm
to shift the sensitivity of clock-based dark matter searches to higher oscillation frequencies while remaining sensitive to a broadband of DM masses \cite{BBDD}. In the dynamic decoupling algorithm, Ramsey pulse sequences with $\pi$ pulses inserted into the free evolution time at random times with a distribution designed to maximize sensitivity to the oscillation frequency range of interest and to dynamically decouple noise from the signal in a quantum system. Reference \cite{BBDD} specifically demonstrates that this approach allows reaching the 10$^{-13}$\,eV mass range without significant sensitivity loss. However, reaching a mass range peak of the Jupiter or Earth halo is beyond present clock architectures and requires different quantum technologies discussed below. We note that other space clock missions aimed at testing general relativity as well as other dark matter searches were recently proposed \cite{FOCOS,ISS}. 

As a result, quantum clocks are sensitive to dark matter that can form Solar halo, but with all there other halos in the Solar system can be formed only by dark matter with larger masses, needing different technologies. 

The limitation of relatively low cadence can be overcome by tuning the laser frequency to a slope of resonance and observing oscillations in the transmitted (or optical rotation) signal as a function of time. If the laser frequency stabilization is configured to be insensitive or differently sensitive to the variation of constants, oscillations of the transition frequency will translate into oscillations of the signal. This idea was implemented with Cs atoms and a polarization-spectroscopy setup \cite{Antypas2019WRESL}, which was further improved to explore the frequency range from near-dc up to 100\,MHz ($4\cdot 10^{-7}$\,eV) with the sensitivity to fractional frequency variations as low as parts in 10$^{18}$ in a part of the explored frequency range \cite{Tretiak2022_RFDMPRL}. 

The measurements \cite{Antypas2019WRESL,Tretiak2022_RFDMPRL} were primarily sensitive to variations of the fine-structure constant $\alpha$ and the mass of the electron. To gain sensitivity to the variation of nuclear mass, the technique was extended to molecular spectroscopy \cite{Oswald:2021vtc} as well as comparisons of atomic transition frequencies with those of a quartz oscillator \cite{Zhang2023_DM_RF}.  

\section{Technical considerations for a spectroscopic search}

The upper frequency limit is determined by several critical factors. First, it is the transition width. As detailed in \cite{Tretiak2022_RFDMPRL} for an atomic transition, the resonance slope is crucial for converting frequency variations into either amplitude modulation or polarization rotation of transmitted light. 
Doppler-free spectroscopy offers a sharp resonance with a cut-off frequency typically below 10\.MHz. In contrast, Doppler-broadened resonances present higher cut-off frequencies but a diminished resonance slope, leading to a trade-off between broader bandwidth and high sensitivity. Given that our primary focus is on Jupiter and the Sun, it is prudent to employ a Doppler-free scheme along with a narrow natural linewidth transition.

Second, the bandwidth of the photodetector defines the analog limitations of the apparatus, thereby constraining the detectable frequency range. In  \cite{Tretiak2022_RFDMPRL}, the photodetector exhibited a cut-off frequency of 100\,MHz ($4\cdot10^{-7}$\,eV).

Another factor is the digitizer sampling frequency, which determines the Nyquist frequency, thereby setting the upper limit of measurable frequencies. Faster digitizers can capture higher frequencies, but this comes at a cost of increased power consumption and reduced data acquisition time due to the system random access memory (RAM) limitations. Therefore, it is important to balance the digitizer speed and RAM capacity to optimize performance.

A major consideration for space-borne missions collecting significant amounts of data is the modest data rate that can be transmitted to Earth, typically below 100\,kbps. This means that the data need to be pre-processed on the spacecraft, ideally, in real time to avoid storage and ensuring continuous operation of the detector. In fact, this is also how it was done in the lab experiments \cite{Tretiak2022_RFDMPRL} using the Fast Fourier Transform (FFT) algorithm.

Given a fixed processor performance, the execution time of the FFT increases proportionally to \(N \log(N)\), where \(N\) is the number of data points. Assuming unlimited RAM and that the data acquisition time equals the FFT execution time, the maximum feasible sampling rate is proportional $N/(N \log(N)) = 1/\log(N)$. 
From these conditions, we can find a balance between data-acquisition time (which determines FFT resolution) and the achievable sampling rate for a given central processing unit performance.

We conduced a benchmark test, in which a 1.2\,GHz Allwinner H3 quad-core ARM Cortex-A7 processor with 512\,MB of RAM was able to perform an FFT of $2^{23}$ samples within 10\,s using only a single core and consuming less than 10\,W of power. This indicates that the system can handle a Fourier transform for 10\,s of data acquisition at a sampling rate of 800\,kS/s with a spectral resolution of 0.1\,Hz. It is expected that more advanced space-qualified electronics will significantly outperform this benchmark, and both spectral resolution and sampling rate could be significantly improved.

For a space apparatus going to Jupiter, an attractive approach is to use a Doppler-free Cs spectroscopy scheme, similar to that described in \cite{Antypas2019WRESL} or \cite{Tretiak2022_RFDMPRL} (Apparatus A). The reasonable amount of acquiring data points per spectrum is the same as in \cite{Tretiak2022_RFDMPRL}, specifically \(2^{28}\), which requires approximately 16\,GB of memory for both data acquisition and processing. With a sampling rate of 100\,kS/s, the acquisition time would be around 45\,minutes, providing a resolution of approximately 0.37\,mHz. This resolution enables detection of dark matter signals within the Jupiter halo, allowing for the lineshape to be resolved with an expected quality factor (Q-factor) on the order of \(10^7\). To achieve this frequency resolution the system has to be provided with a reference clock with accuracy better then $10^{-9}$ over the data acquisition time. These requirements are well within modern capabilities.

Of primary interest for us in this work are the relatively low frequencies ($< 20$\,kHz corresponding to UBDM-particle mass of $<10^{-10}$\,eV),  where the largest enhancements due to Sun and Jupiter are expected (see Figs.\,\ref{fig:GAdensity} and \ref{fig:sensitivity}). This regime is complementary to that in 
 \cite{Tretiak2022_RFDMPRL}, where the measurement of frequency modulation of cesium was limited from the bottom by 20\,kHz ($8\cdot10^{-11}$\,eV). The atomic transition frequency was compared to that of a laser stabilized in its own cavity or a Fabry-P\'erot reference cavity. The lower limit of the frequency span was defined by the mechanical wave propagation in the spacer between the cavity mirrors. Below this frequency, the fine-structure constant and the electron-mass variations impact not only the atomic transition but also the size of the cavity. This must be considered during result interpretation \cite{Antypas_2021_QST}.
 
In the range of 0 to 20\,kHz, numerous sources of mechanical vibrations can interfere with the measurement. To operate effectively within this range, the optical cavity and the entire system must be isolated from environmental influences. This problem should be tractable in space via a combination of isolation and vetoing data recorded during the operation of noisy equipment such as the operation of orbit-correction systems.

\section{Proposed space-borne experiments}

We propose to incorporate detectors of UBDM-induced fast apparent oscillation of fundamental constants in the future space missions in the Solar System, in particular particular, the Jupiter and solar ones.
A dedicated Cs atomic spectrometer as described above is a good candidate, although other atomic and molecular systems should also be considered, with molecular systems particularly sensitive to variation of nuclear mass.



Because this kind of atomic or molecular spectrometer is designed for a search for a narrow ($\delta f/f \ll 10^{-6}$) 
at frequencies above the typical $1/f$-noise ``knee'', the setups tend to be technically simpler than high-precision atomic clocks and offer straightforward paths towards miniaturization and power consumption dominated by the laser source (with output power in the milliwatt range). Moreover, the basic hardware is not much different from what is used for laser stabilization in conjunction with atomic clocks, magnetometers and other atom-based sensors, including cold atom based setups. 

It is important to note that, in addition to the scalar couplings leading to apparent oscillation of fundamental constants accessible with the spectrometers discussed above, other couplings will be accessible with other types of sensors. For example, atomic magnetometers are sensitive to gradients of pseudoscalar fields as well as spin-one fields (such as the dark or hidden photon). Thus comprehensive direct halo searches with multisensor space missions can be envisioned.

\section{Conclusions}
\label{sec:conclusion}
In this work, we show that space missions with various quantum technologies enable density-enhanced searches for a broad range of dark matter masses maximizing the discovery potential. Atomic and molecular spectroscopy technologies described in this work enable sufficiently compact dimensions and modest power requirements so that they can be incorporated in future planetary missions.  

\section*{Acknowledgments}

The authors acknowledge helpful discussions with Gilad Perez. This research was supported in part by the Deutsche Forschungsgemeinschaft (DFG, German Research Foundation) Project ID 390831469: EXC 2118 (PRISMA+ Cluster of Excellence), by the COST Action within the project COSMIC WISPers (Grant No. CA21106), and the Munich Institute for Astro-, Particle and BioPhysics (MIAPbP), which is funded by the DFG under Germany´s Excellence Strategy – EXC-2094 – 390783311, and by the National Science Foundation grants PHY-2309254 and Q-SEnSE Quantum Leap Challenge Institute (Grant Number OMA-2016244).
The work of J.E.~is supported by the Swedish Research Council (VR) under grants 2018-03641 and 2019-02337.

\appendix

\section{Direct capture as formation mechanism for Jupiter-bound halo}
\label{app:capture}

In the mechanism of Ref.\,\cite{Budker2023_Formation}, the self-interactions in Eq.\,\eqref{eq:Lagrangian} give rise to $2\to 2$ scattering of $\phi$ particles in the gravitational field of the central body.
The formation of a dense gravitational atom by this process is of astrophysical interest, given two conditions. First, in order for the $1/r$ approximation of the gravitational potential to be a reasonable approximation, we require 
\begin{equation} \label{eq:cond1}
    R_\star = (m_\phi \alpha)^{-1} 
            \gtrsim R \,.
\end{equation}
Second, in order for rapid capture on astrophysical timescales, significant gravitational focusing of the field is necessary, which is active only when the de Broglie wavelength of the incoming DM is larger than $R_\star$~\cite{Kim:2021yyo}, or equivalently~\cite{Budker2023_Formation}
\begin{equation} \label{eq:cond2}
    M \gtrsim \frac{v_{\rm dm}}{2\pi G m_\phi}\,.
\end{equation} 

\begin{figure}[t!]
    \centering
    \includegraphics[scale=0.57] {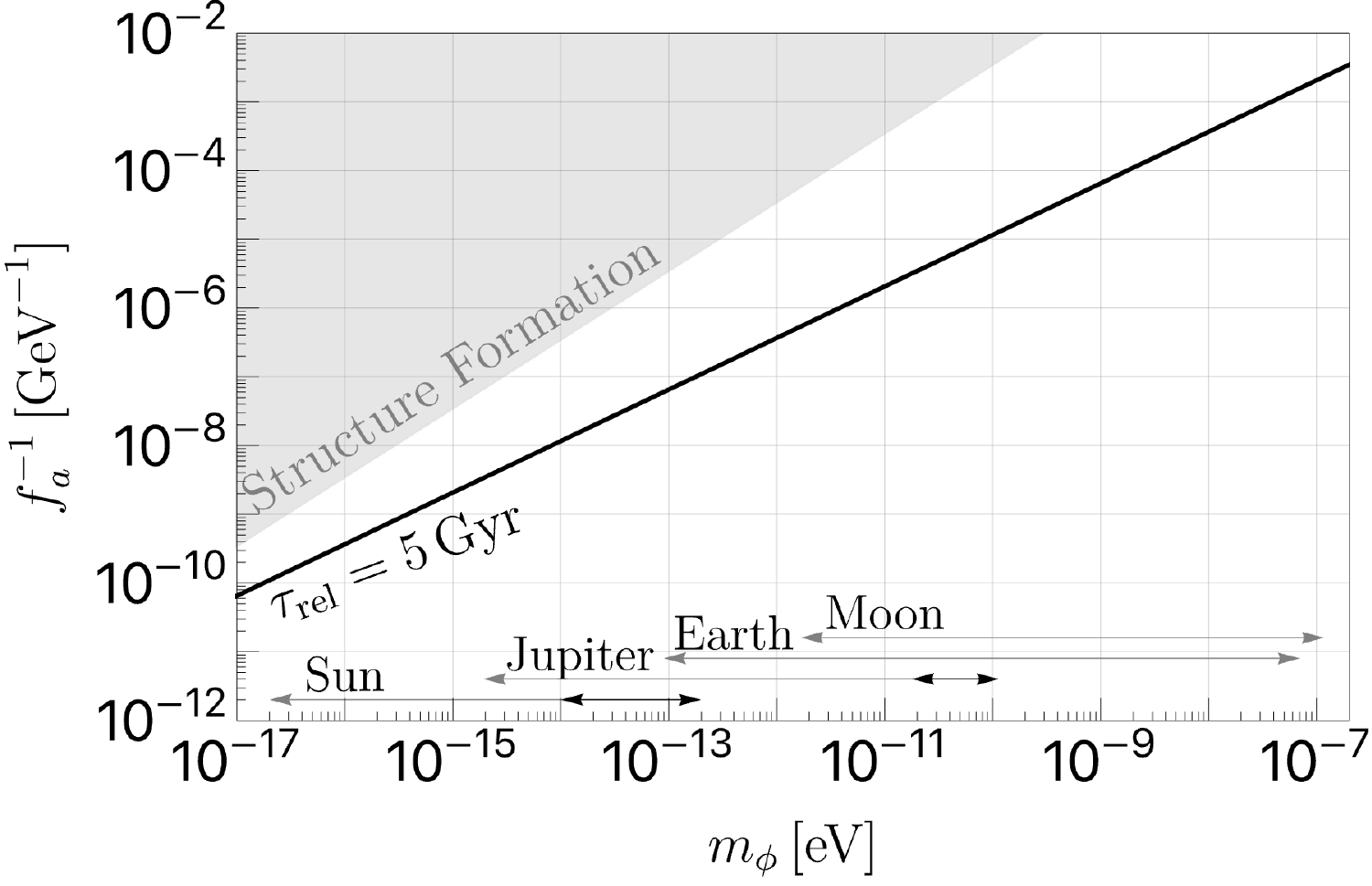}
    \caption{The line corresponding to relaxation time $\tau_{\rm rel}=5\,{\rm Gyr}$ in the space of decay constant $f_a$ and UBDM mass $m_\phi$. For comparison, the arrows indicate the region of interest for a gravitational atom bound to the objects we consider in this work (see text for details). The gray shaded region is constrained by large-scale structure surveys.
    }
    \label{fig:structure_bounds}
\end{figure}

The self-interaction strength is important for setting the timescale for the formation of the gravitational atom, and is parameterized by the dimensionless coupling $-m_\phi^2/f_a^2$. We summarize the parameter space of $m_\phi$ and $f_a$ in Fig.\,\ref{fig:structure_bounds}, highlighting several relevant features. 
The blue solid line shows the relation $\tau_{\rm rel}=5\,{\rm Gyr}$, where the relaxation time of the ultralight scalar field is given by~\cite{Levkov:2018kau,Chen:2020cef,Kirkpatrick:2020fwd,Chen:2021oot,Kirkpatrick:2021wwz,Dmitriev:2023ipv,Jain:2023tsr}
\begin{equation}   
    \tau_{\rm rel} \equiv \frac{64 m_\phi^3 f_a^4 v_{\rm dm}^2}{\rho_{\rm local}^2}\,.
\end{equation}
The value of $5\,{\rm Gyr}$ is chosen as a proxy for the lifetime of the solar system.
On the other hand, the gray shaded region indicates the constraint from large-scale structure; in this range, sufficiently-strong self-interactions can enhance or suppress the matter power spectrum on scales larger than Mpc, in conflict with observation~\cite{Arvanitaki:2014faa,Fan:2016rda,Cembranos:2018ulm}.
Importantly, a relaxation time below $\sim$few Gyr is possible in the allowed region of parameters.

For illustration, the horizontal arrows labeled by central objects (e.g. ``Sun'') show the range of $m_\phi$ in which a bound state can exist and have density greater than the background value $\rho_{\rm local}$ (c.f. Fig.\,\ref{fig:GAdensity}); the gray arrows show the full range, whereas the black arrows show the range satisfying the conditions in Eq.\,(\ref{eq:cond1}-\ref{eq:cond2}). 

\begin{figure}[b!]
    \centering
    \includegraphics[scale=0.57] {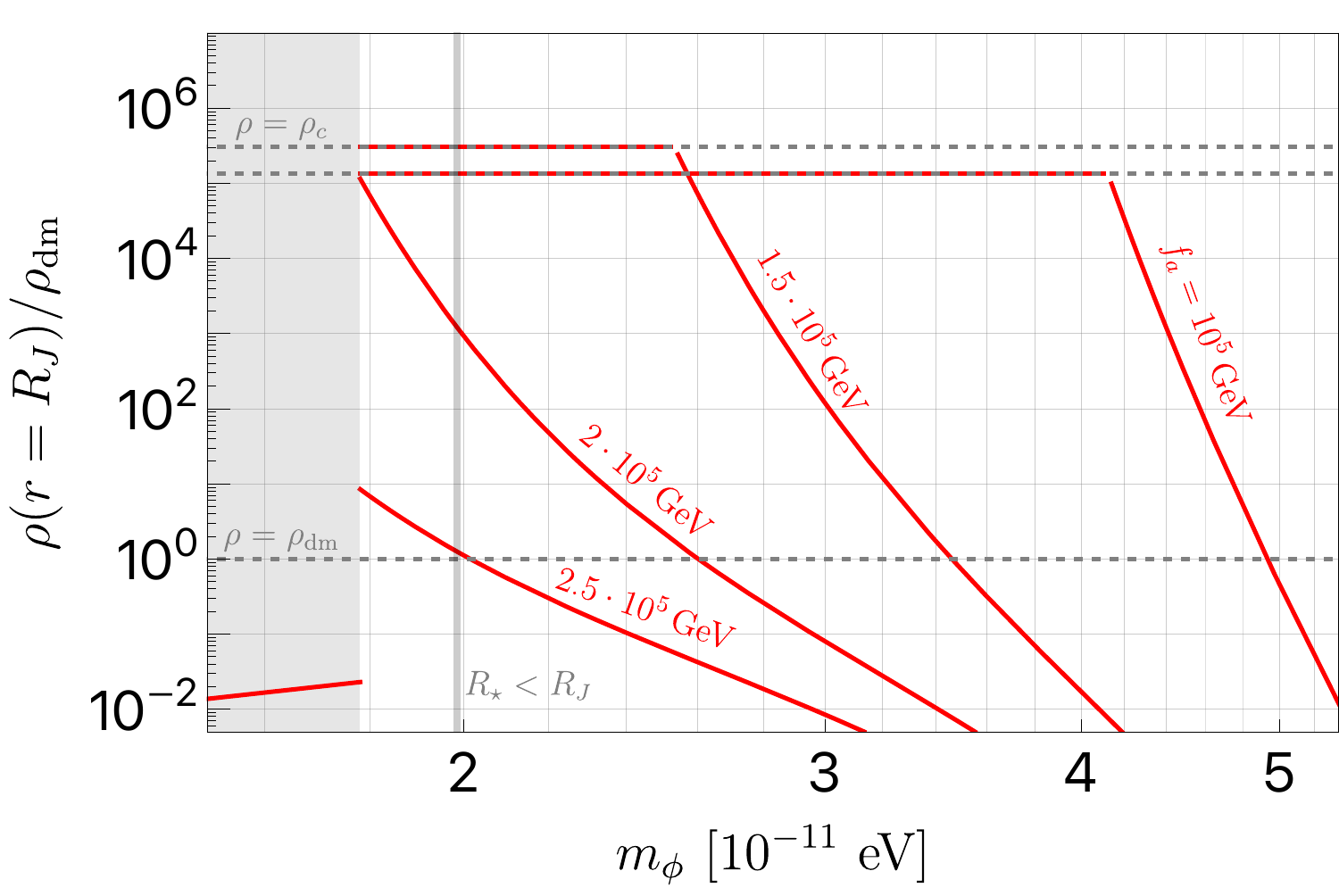}
    \caption{The density bound to Jupiter after $4.5\,{\rm Gyr}$ for DM capture via self-interactions~\cite{Budker2023_Formation}, assuming a decay constant $f_a$ as labeled. The critical density $\rho_c$ and background density $\rho_{\rm dm}$ are given by the dashed lines. For $m_\phi \lesssim 10^{-11}\,{\rm eV}$ the growth is slow due to kinematic suppression (gray shaded region) and the density saturates to the lower red line (see  \cite{Budker2023_Formation} for details). 
    }
    \label{fig:GAdensity_Jupiter}
\end{figure}

For the Sun, these conditions are simultaneously satisfied for $10^{-13}\,{\rm eV} \lesssim m_\phi \lesssim 10^{-14}\,{\rm eV}$~\cite{Budker2023_Formation}, whereas for the Earth and the moon they are not satisfied for any $m_\phi$. In this Appendix we explore the range of applicability for a gravitational atom surrounding Jupiter. By computing the total captured density over $4.5\,{\rm Gyr}$, Jupiter's approximate lifetime, we illustrate the maximum Jupiter-bound halo density by the red lines in Fig.\,\ref{fig:GAdensity_Jupiter}. These lines are labeled by the value of $f_a$, and are cut off by the dotted lines where $\rho_\star = \rho_c$ in Eq.\,\eqref{eq:rhoc}.

\section{Constraints on the DM density around the solar system}
\label{app:constraints}

Here we review the most relevant observational constraints of interest for this work. These constraints depend importantly on the scale of the overdensity, whether it be $\sim$AU as for solar-bound DM or much smaller for planet-bound or moon-bound DM.

On AU scales, precise ephemerides of solar system objects 
lead to constraints on dark matter centered around the Sun. These constraints are at the level of $\rho\lesssim \{10^7,10^5,10^4\}\rho_{\rm local}$ at the position of Mercury, Earth, and Saturn (respectively), with similar limits also at the positions of Venus, Mars, and Jupiter~\cite{Pitjev:2013sfa}. The overdensities for the most distant planets (Uranus and Neptune) have been constrained similarly at the level of $\rho\lesssim 10^7\rho_{\rm local}$~\cite{Gron:1995rn, Anderson:1995dw}. Recently it was shown that precision asteroid tracking can set a competitive constraint on AU scales as well~\cite{Tsai:2022jnv}, providing the possibility of improvement in the near future. 

For DM centered around the Earth, the total bound DM mass $M_\star$ can be constrained by comparing the orbit of the moon with objects in low-Earth orbit, e.g. the LAGEOS satellite.
A total DM mass $M_\star \gtrsim 4\cdot10^{-9} M_\oplus$ would modify the relative orbits of the moon and LAGEOS satellite~\cite{Adler:2008ky}. For comparison, at the orbital radius of LAGEOS (of the moon) this corresponds to a maximum density of order $\rho\simeq 10^{15}\rho_{\rm local}$, obtained for $m_\phi\simeq 10^{-9}\,{\rm eV}$ ($\rho\simeq 10^{10}\rho_{\rm local}$ for $m_\phi\simeq 10^{-10}\,{\rm eV}$). 
This constraint can be approximately extended to a moon-centered halo, as $\mathcal{O}(1/2)$ of the bound DM mass would lie between these orbits, at least for $R_\star \lesssim R_{\oplus-{\rm moon}}$, the Earth-moon distance.\footnote{The fraction is smaller than $1/2$ for $R_\star > R_{\oplus-{\rm moon}}$, making this estimate conservative.} 

Moons and satellites around other planets (besides Earth) should give rise to analogous constraints, though to our knowledge no precise analysis of this kind has so far been performed. In particular, precise tracking of Jupiter's largest moons should give rise to relevant constraints. Beyond naturally-occurring satellites, the Juno mission~\cite{JUNO_mission:2017abc} would be a particularly promising for probing the DM density around Jupiter, a topic we hope to return to in the near future.

%

\end{document}